%% file: main.tex
\documentclass[letterpaper, 10 pt, conference]{ieeeconf}  

\IEEEoverridecommandlockouts                              

\usepackage{graphics} 
\usepackage{graphicx} 
\usepackage{epsfig} 
\IEEEoverridecommandlockouts        
\usepackage{times} 
\usepackage{amsmath} 
\usepackage{amssymb}  
\usepackage{array}
\usepackage{cite}
\usepackage{color}
\usepackage{comment}
\usepackage{epstopdf}
\usepackage{subfigure}
\usepackage{float} 
\usepackage{verbatim}
\usepackage{tabularx}
\usepackage{booktabs} 
\usepackage{multirow}
\usepackage{diagbox}
\usepackage{siunitx}
\usepackage{mathrsfs}

\input{defs}
\title{\LARGE \bf Information Flow Topology in Mixed Traffic: A  Comparative Study between ``Looking Ahead" and ``Looking Behind"}

\author{Shuai Li, Haotian Zheng, Jiawei Wang, Chaoyi Chen, Qing Xu, Jianqiang Wang and Keqiang Li
\thanks{This work is supported by National Key R\&D Program of China with 2021YFB1600402, National Natural Science Foundation of China, Science Fund for Creative Research Groups with 52221005, and Tsinghua-Toyota Joint Research Institute Cross-discipline Program. Corresponding author: Jiawei Wang.}
\thanks{S.~Li, H.~Zheng, J.~Wang, C.~Chen, Q.~Xu, J. Q.~Wang and K.~Li are with the School of Vehicle and Mobility, Tsinghua University, Beijing, China. (\{li-s21, zhenght21, wang-jw18\}@mails.tsinghua.edu.cn, chency2023@mail.tsinghua.edu.cn, \{qingxu, wjqlws, likq\}@tsinghua.edu.cn).}
}

\begin{document}

\maketitle
\thispagestyle{empty}
\pagestyle{empty}

\begin{abstract}

The emergence of connected and automated vehicles (CAVs) promises smoother traffic flow. In mixed traffic where human-driven vehicles (HDVs) also exist, existing research mostly focuses on ``looking ahead" (\emph{i.e.}, the CAVs receive information from preceding vehicles) strategies for CAVs, while recent work reveals that ``looking behind" (\emph{i.e.}, the CAVs receive information from their rear vehicles) strategies might provide more possibilities for CAV longitudinal control. This paper presents a comparative study between these two types of information flow topology (IFT) from the string stability perspective, with the role of maximum platoon size (MPS) also under investigation. 
Precisely, we provide a dynamical modeling framework for the mixed platoon under the multi-predecessor-following (MPF) topology and the multi-successor-leading (MSL) topology. Then, a unified method for string stability analysis is presented, with explicit consideration of both IFT and MPS. Numerical results suggest that MSL (``looking behind") outperforms MPF (``looking ahead") in mitigating traffic perturbations. In addition, increasing MPS could further improve string stability of mixed traffic flow.

\end{abstract}

\section{Introduction}
With recent advancements in vehicle-to-everything (V2X) communication and vehicle control, the emergence of connected and automated vehicles (CAVs) promises to address the persistent challenges in road traffic, including frequent accidents, traffic congestion, and heightened energy consumption~\cite{Wang2019a,Hu2022}. In particular, enabling CAVs to improve traffic stability and mitigate traffic waves has attracted increasing attention in traffic engineering~\cite{guanetti2018control,Hu2017}. In practice, nevertheless, the complete marketization of CAVs necessitates a gradual progression, resulting in a long-term phase of mixed traffic where both CAVs and human-driven vehicles (HDVs) coexist~\cite{zheng2020smoothing,stern2018dissipation}. Previous simulation-based investigations concerning mixed traffic have demonstrated that augmenting the penetration rate of CAVs can contribute to a traffic improvement, which might, however, be limited in low penetration rates~\cite{Milanes2014,talebpour2016influence}.
 
Platooning is a typical CAV technology to improve traffic characteristics, with string stability recognized as a prominent index~\cite{feng2019string}. String stability focuses on whether oscillations of the downstream vehicles are amplified as they propagate upstream in the traffic flow. Traffic waves usually occur if string stability is not guaranteed. Researchers have endeavored to enhance the string stability of mixed traffic by employing pure CAV platoons, revealing that the information flow topology (IFT) (\emph{i.e.}, the information exchange relationship between vehicles in the platoon) plays a critical role~\cite{Zheng2015a}. Both theoretical analysis and traffic simulations have been conducted to enhance string stability by choosing an appropriate and feasible IFT for platoons~\cite{wu2019distributed,feng2020robust}. 
It is worth noting that these investigations have focused on the IFT for pure CAV platoons (\emph{i.e.}, the composition of platoon includes only the CAVs), rather than the mixed platoon (\emph{i.e.}, the platoon consists of CAVs and HDVs), which is more common in mixed traffic flow~\cite{chen2021mixed,li2022cooperative}. 

\begin{figure*}[!ht]
    \centering
    \vspace{0.2cm}
    \subfigure[Multi-predecessor-following (MPF) topology, also known as ``looking ahead"]{\includegraphics[width=16.5cm]{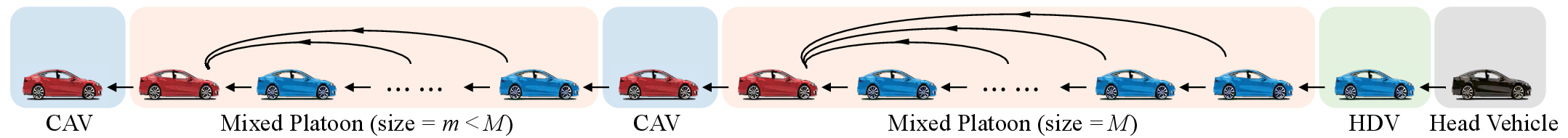}}\\
    \vspace{-0.2cm}
    \subfigure[Multi-successor-leading (MSL) topology, also known as ``looking behind"]{\includegraphics[width=16.4cm]{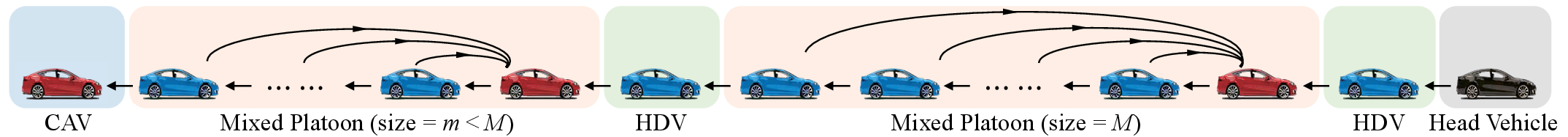}}
    \vspace{-0.2cm}
    \caption{Schematic for typical information flow topologies in mixed traffic flow. The black, red and blue vehicles represent head vehicle, CAVs and HDVs, respectively. The gray, orange, blue, and green boxes represent the head vehicle, mixed platoons, independent CAVs, and independent HDVs.}
    \label{fig:1}
    \vspace{-0.4cm}
\end{figure*}

Connected cruise control (CCC) is a particular concept for mixed platoons, which generalizes the multi-predecessor-following (MPF) topology from pure CAV platoons to the mixed traffic scenario~\cite{orosz2016connected}. Particularly, CCC utilizes information from multiple preceding HDVs to design the control inputs for the CAV located at the tail of the mixed platoon. Along this direction, several topics have been discussed, such as the impact of the optimal controller's cost function~\cite{Jin2016} and compensation for communication delay~\cite{molnar2017application}. Informally, this kind of controller under the MPF topology can be regarded as ``looking ahead" strategies; see Fig.~\ref{fig:1}(a) for demonstration. As an extension of CCC to more general scenarios, the recent concept of leading cruise control (LCC) introduces ``looking behind" topologies into CAV control in mixed traffic, where the CAV not only needs to follow the HDVs ahead, but also attempts to lead the motion of the HDVs behind~\cite{wang2022leading}. Formally, one particular kind of ``looking behind" topology can be named as multi-successor-leading (MSL); see Fig.~\ref{fig:1}(b) for demonstration. Several studies have demonstrated the potential of MSL control strategy in improving traffic stability and efficiency~\cite{wang2022deep,wang2022distributed}.

To the best of our knowledge, however, an explicit comparison between ``looking ahead" and ``looking behind" strategies is still lacking. A very recent work in~\cite{ruan2022impacts} presents a comparison between MPF and other ``looking ahead" topologies, but the case of ``looking behind" is beyond consideration. To address this problem, this paper provides a unified method to analyze their influence on string stability of mixed traffic flow, and aims to answer the open question of which kind of strategy might lead to better traffic-level improvement. In addition, we also incorporate the effects of maximum platoon size (MPS) (\emph{i.e.}, the maximum number of vehicles that can be allowed in the platoon), which has been recently recognized as a critical factor in traffic flow stability~\cite{yao2023analysis}. Due to the practical constraints of communication bandwidth and distance, MPS and IFT have a coupled relationship, where the former directly limits the latter's structure. Existing research on MPS mostly focuses on pure CAV platoons (see, \eg,~\cite{yao2023analysis,Zhou2021}), and its role in mixed platoons remains unclear. 
To fill the gap, this paper delves into the coupled influence of IFT and MPS on the string stability of mixed traffic within mixed platoons. Precisely, the contributions are as follows: 

\begin{itemize}
  \item We present a unified method to analyze the string stability of mixed traffic flow. Compared with the standard tool in~\cite{feng2019string}, the role of IFT and MPS are both under explicit consideration. 
  
  \item Based on the proposed string stability analysis framework, we investigate the impact of different types of IFT. Unlike existing research in~\cite{ruan2022impacts}, which solely focuses on the ``looking ahead" topologies, our study presents a first comparison between ``looking ahead" and ``looking behind" strategies. 

  \item  We also provide a comprehensive analysis of MPS in mixed traffic. Existing work mostly considers pure CAV platoon scenarios; see, \emph{e.g.},~\cite{yao2023analysis,Zhou2021}. In this work, instead, we focus on the size configuration of the mixed platoon and shed the first insight into optimizing the MPS in mixed traffic flow.


  

\end{itemize}

The remainder of this paper is organized as follows. Section~\ref{Sec:2} presents the problem statement and the modeling for the mixed traffic system. Section~\ref{Sec:3} presents the string stability analysis and comparisons for different IFT and MPS, and the nonlinear traffic simulations are shown in Section~\ref{Sec:4}. Finally, Section~\ref{Sec:5} concludes this paper.

\section{Problem Statement and System Modeling}
\label{Sec:2}

In this section, we present the problem statement and the dynamical modeling for the mixed traffic system.
\subsection{Problem Statement}

In this paper, we focus on the scenario of mixed traffic on a single-lane straight road, where CAVs and HDVs are randomly distributed, as shown in Fig.~\ref{fig:1}. The head vehicle is indexed as $0$ and other vehicles are indexed as $1,2,\ldots,N$ against the moving direction. Two typical IFT types are under consideration: 1) MPF, where the CAVs gather information from multiple preceding vehicles, as illustrated in Fig.\ref{fig:1}(a), and 2) MSL, where the CAVs receive information from their rear vehicles, as shown in Fig.\ref{fig:1}(b). To ensure practical applicability, the CAVs in MSL also need to consider the motion of the vehicle immediately ahead for basic car-following operation and collision avoidance. Note that the MPF topology is essentially a CCC-type framework~\cite{orosz2016connected}, while the MSL topology studied in this paper is a special case of LCC, called car-following LCC~\cite{wang2022leading}.

Furthermore, we define mixed platoon in this paper as a series of vehicles of one single CAV and its neighbouring HDVs. Then, the mixed traffic can be divided by the location of the CAVs into a sequence of mixed platoon subsystems. The MPS is denoted as $M$, and when the total number of the vehicles in a mixed platoon reaches $M$, the CAV only acquires motion information from the HDVs within the platoon, excluding the vehicle motion information outside the platoon. Define the actual size of the mixed platoon as $m$, and we have $M \geq m$.  
Take MPF in Fig.~\ref{fig:1}(a) as an example. If there are no more than $M-1$ HDVs in front of a certain CAV, then that CAV and the HDVs are grouped into a mixed platoon; otherwise, the $M-1$ HDVs immediately ahead are incorporated into the mixed platoon, while the other HDVs are regarded to be not connected to any CAV; see the HDVs in green boxes in Fig.~\ref{fig:1}.  In addition, those CAVs with a CAV neighbour are regarded as independent vehicles, rather than being grouped into a mixed platoon; see the CAVs in blue boxes in Fig.~\ref{fig:1}.


\subsection{Modeling of the Mixed Platoon System}

We proceed to present the modeling for the mixed platoon system. 
According to existing studies, 
typical car-following models for HDVs, such as the intelligent driver model (IDM)~\cite{Treiber2000} and optimal velocity model (OVM)~\cite{Bando1998}, have the following general form
~\cite{Jin2016,wang2022leading}:
\begin{equation}
\label{deqn_1}
\dot{v}_i(t)=f_i\left(s_i(t), \dot{s}_i(t), v_i(t)\right),
\end{equation}
where $\dot{v}_i(t)$ is the acceleration of vehicle $i$, $s_i(t)$ is the space between vehicle $i-1$ (the preceding vehicle) and vehicle $i$, $\dot{s}_i(t)=v_{i-1}(t)-v_i(t)$ is the velocity difference, and $v_i(t)$ is the velocity of vehicle $i$. Note that in this paper, for independent HDVs and CAVs, we also assume that their model follows the form of~\eqref{deqn_1}.


During regular driving, individual vehicles usually encounter minor disturbances, prompting our analysis to focus on the near equilibrium state. In the equilibrium state, each vehicle moves with the same equilibrium velocity $v^{*}$ and the desired space $s^{*}$, \emph{i.e.}, $v_i(t)=v^{*}$, $s_i(t)=s^{*}$, for $i=1,2,...,n$. Thus, every HDV in the equilibrium
state must satisfy the following equilibrium equation:
\begin{equation}
\label{deqn_2}
f_i(s^{*},0,v^{*})=0, i=1,2,...,n.
\end{equation}

Denote  $\tilde{s}_i(t)={s}_i(t)-s^{*}$  and $\tilde{v}_i(t)={v}_i(t)-v^{*}$ as the deviation of the space and velocity from the equilibrium state, and we define the error state for each vehicle as: 
\begin{equation}
\label{deqn_3}
x_i(t) = \left [ \tilde{s}_i(t),\tilde{v}_i(t) \right ]^{\top} .
\end{equation}

Using the first-order Taylor expansion for~\eqref{deqn_1}, we obtain the linearized model for each HDV:
\begin{equation}
\label{deqn_4}
\dot{\tilde{v}}_{i}(t)
=\dot{f}_{i}^{s} \tilde{s}_{i}(t)-\left(\dot{f}_{i}^{\dot{s}}-\dot{f}_{i}^{v}\right) \tilde{v}_{i}(t)+\dot{f}_{i}^{\dot{s}} \tilde{v}_{i-1}(t),
\end{equation}
where $\dot{f}_{i}^{s}$, $\dot{f}_{i}^{\dot{s}}$, and $\dot{f}_{i}^{v}$ are the partial derivatives of~\eqref{deqn_1} in the equilibrium state concerning the space, velocity difference, and velocity, respectively. Let $\alpha_{1}=\dot{f}_{i}^{s}$ , $\alpha_{2}=\dot{f}_{i}^{\dot{s}}-\dot{f}_{i}^{v}$, and $\alpha_{3}=\dot{f}_{i}^{\dot{s}}$, and then the linearized HDV model can be obtained as follows:
\begin{equation}
\label{deqn_5}	
\begin{cases}
\dot{\tilde{s}}_{i}(t)=\tilde{v}_{i-1}(t)-\tilde{v}_{i}(t), \\
\dot{\tilde{v}}_{i}(t)=\alpha_{1} \tilde{s}_{i}(t)-\alpha_{2} \tilde{v}_{i}(t)+\alpha_{3} \tilde{v}_{i-1}(t). 
\end{cases}
\end{equation}

For the MPF topology, the acceleration signal is used as the control CAVs' input $u_i(t)$, and we have
\begin{equation}
\label{deqn_6}	
\begin{cases}
\dot{\tilde{s}}_{i}(t)=\tilde{v}_{i-1}(t)-\tilde{v}_{i}(t), \\
\dot{\tilde{v}}_{i}(t)=u_i(t).
\end{cases}
\end{equation}

For the MSL topology, 
we assume that the CAV adopts the linearized HDV model to follow the vehicle immediately ahead, and meanwhile incorporates a control input denoted as $u_i(t)$ to regulate the CAV's behavior based on the vehicles behind. Precisely, its dynamics is given by:
\begin{equation}
\label{deqn_7}	
\begin{cases}
\dot{\tilde{s}}_{i}(t)=\tilde{v}_{i-1}(t)-\tilde{v}_{i}(t), \\
\dot{\tilde{v}}_{i}(t)=\alpha_{1} \tilde{s}_{i}(t)-\alpha_{2} \tilde{v}_{i}(t)+\alpha_{3} \tilde{v}_{i-1}(t)+u_i(t).
\end{cases}
\end{equation}

Then, we lump the error states of all the involved vehicles as the aggregate state of the mixed platoon. Specifically, the state vector of the mixed platoon system  under the MPF topology is defined as:
\begin{equation}
\label{deqn_8}	
x(t)=\left[x_{i-m+1}(t), \ldots, x_{i-1}(t), x_{i}(t)\right]^\top,
\end{equation}
and the state vector of the mixed platoon system under the MSL topology is: 
\begin{equation}
\label{deqn_9}	
x(t)=\left[x_{i}(t), x_{i+1}(t), \ldots, x_{i+m-1}(t)\right]^\top.
\end{equation}

Then, a linearized state-space equation for mixed platoon is obtained as	
\begin{equation}
\label{deqn_10}	
\begin{cases}
\dot{x}(t)=A x(t)+B u_i(t)+H \tilde{v}_{p}(t), \\
y(t)=C x(t),
\end{cases}
\end{equation}
where $A \in \mathbb{R}^{2m \times 2m}$  is the system matrix, $B \in \mathbb{R}^{2m \times 1}$ is the control input matrix, $H \in \mathbb{R}^{2m \times 1}$ is the disturbance input matrix, $C \in \mathbb{R}^{1 \times 2m}$ is the output matrix, $y(t)$ is the output of the mixed platoon system, $u_i(t)$ is the control input of the CAV, and $\tilde{v}_{p}(t)$ is the velocity deviation of the closest preceding vehicle of the mixed platoon. For MPF topology, the $A$, $B$, and $H$ are given as follows:
\begin{equation}
\label{deqn_11}	
\left[\begin{array}{ccccc}
A_{1} & & & & \\
A_{2} & \ddots & & & \\
& \ddots & \ddots & & \\
& & A_{2} & A_{1} & \\
& & & A_{4} & A_{3}
\end{array}\right], \;\left[\begin{array}{c}
B_{1} \\
\vdots \\
\vdots \\
B_{1} \\
B_{2} 
\end{array}\right], \;\left[\begin{array}{c}
H_{1} \\
H_{2} \\
\vdots \\
\vdots \\
H_{2}
\end{array}\right].
\end{equation}
For mixed platoon system under MSL topology, the $A$, $B$, and $H$ are given by:
\begin{equation}
\label{deqn_12}	
\left[\begin{array}{ccccc}
    A_{1} & & & & \\
    A_{2} & \ddots & & & \\
    & \ddots & \ddots & & \\
    & & A_{2} & A_{1} & \\
    & & & A_{2} & A_{1}
\end{array}\right], \;\left[\begin{array}{c}
    B_{2} \\
    B_{1} \\
    \vdots \\
    \vdots \\
    B_{1}
\end{array}\right], \; \left[\begin{array}{c}
    H_{1} \\
    H_{2} \\
    \vdots \\
    \vdots \\
    H_{2}
\end{array}\right].
\end{equation}
For both MPF topology and MSL topology, the output matrix  $C$ is the same, which is expressed by:
\begin{equation}
\label{deqn_13}	
\left[\begin{array}{lllll}
C_{1} & \cdots & \cdots & C_{1} & C_{2}
\end{array}\right],
\end{equation}	
Each sub-block matrix in~\eqref{deqn_11}-\eqref{deqn_13} is defined as follows:
\begin{equation}
\label{deqn_14}		
\begin{array}{l}
    A_{1}=\left[\begin{array}{cc}
        0 & -1 \\
        \alpha_{1} & -\alpha_{2}
    \end{array}\right],\;A_{2}=\left[\begin{array}{cc}
        0 & 1 \\
        0 & \alpha_{3}
    \end{array}\right],\\
    A_{3}=\left[\begin{array}{cc}
        0 & -1 \\
        0 & 0
    \end{array}\right],\;A_{4}=\left[\begin{array}{ll}
        0 & 1 \\
        0 & 0
    \end{array}\right],\\
    B_{1}=\left[\begin{array}{l}
        0 \\
        0
    \end{array}\right],\;
    B_{2}=\left[\begin{array}{l}
        0 \\
        1
    \end{array}\right],\;
    H_{1}=\left[\begin{array}{c}
        1 \\
        \alpha_{3}
    \end{array}\right],\\
    H_{2}=\left[\begin{array}{l}
        0 \\
        0
    \end{array}\right],\;
    C_{1}=\left[\begin{array}{ll}
        0 & 0
    \end{array}\right],\; C_{2}=\left[\begin{array}{ll}
        0 & 1
    \end{array}\right].
\end{array}
\end{equation}

\section{String Stability Analysis}
\label{Sec:3}

In this section, we propose a unified method for string stability analysis of mixed traffic  under different topologies, and show the numerical results utilizing specific control strategies.

\subsection{Unified Method}

The traffic flow is called string stability if the oscillations of downstream vehicles of all excitation frequencies do not get amplified by upstream vehicles; otherwise,  the traffic flow is considered as being string unstable. In this study, we employed the approach of transfer function theory  to investigate the string stability of mixed traffic~\cite{feng2019string,talebpour2016influence}. The criterion for string stability of mixed traffic under the MPF and MSL can be expressed as follows:
\begin{equation}
\label{Eq:StringStability}	
\begin{array}{c}
\| G_{M}(\mathscr{S})^{N p_{(\text{size}=M)}} \times \prod_{m=2}^{M-1} G_{m}(\mathscr{S})^{N p_{(\text{size}=m)}} \\
\times G_{\text{CAV}}(\mathscr{S})^{N p_{(\text{CAV})}} \times G_{\text{HDV}}(\mathscr{S})^{N p_{(\text{HDV})}} \|_{\infty} \leq 1,
\end{array}
\end{equation}
where $\mathscr{S}$ is the complex frequency symbol generated by the Laplace transform, $G_{M}(\mathscr{S})$ is the transfer function when the mixed platoon size reaches the $ M $, $G_{m}(\mathscr{S})$ is the transfer function when the mixed platoon size is $ m $ with $m<M$, $G_{\text{CAV}}(\mathscr{S})$ is the transfer function of the independent CAV model, $G_{\text{HDV}}(\mathscr{S})$ is the transfer function of the independent HDV model. We also denote $p_{(\text{size}=M)}$, $p_{(\text{size}=m)}$, $p_{(\text{CAV})}$, $p_{(\text{HDV})}$ as the probability of different transfer functions, which can be calculated as follows:
\begin{equation}
\label{deqn_16}
\begin{cases}
p_{(\text {size}=M)}&=p \times(1-p)^{M-1}, \\
p_{(\text {size}=m)}&=p^{2} \times(1-p)^{m-1},	\\
p_{(\text {CAV})}&=p^2,	\\
p_{(\text {HDV})}&=(1-p)^M,
\end{cases}
\end{equation}
where $p$ denotes the CAVs' penetration rate in mixed traffic.



\subsection{Specific Car-Following and Control Strategies}

For numerical analysis of~\eqref{Eq:StringStability}, we proceed to select the specific car-following models or control strategies for the three types of systems in mixed traffic flow, including independent HDVs, independent CAVs, and mixed platoons, as shown in Fig.~\ref{fig:1}.

\vspace{0.2em}
\noindent
\textbf{Independent HDVs:}
For the independent HDVs and the HDVs involved in the mixed platoon, we consider the OVM model~\cite{Jin2016}: 
\begin{equation}
\label{deqn_17}
\dot{v}_{i}(t)=\alpha\left(V\left(s_{i}(t)\right)-v_{i}(t)\right)+\beta\left(v_{i-1}(t)-v_{i}(t)\right),
\end{equation}
where 
$\alpha>0$ denotes the driver's sensitivity parameter to the difference between the desired velocity and the actual velocity, and $\beta>0$ denotes the driver's sensitivity parameter to the velocity difference between the vehicle and its preceding vehicle. The 
$V\left(s_{i}(t)\right)$ function denotes the desired velocity at the currently space $ s_{i}(t) $, which is usually expressed as:
\begin{equation}
\label{deqn_18}
V\left(s_{i}(t)\right)=
\begin{cases}
0 & \left(s_{i}(t) \leq s_{\min }\right) \\
f_{v}\left(s_{i}(t)\right) & \left(s_{\min }<s_{i}(t)<s_{\max }\right) \\
v_{\max } & \left(s_{i}(t) \geq s_{\max }\right)
\end{cases},
\end{equation}
where $ s_{\min } $ denotes the minimum car-following space, $ s_{\max } $ denotes the maximum car-following space that can generate the car-following behavior, and $f_{v}\left(s_{i}(t)\right) $ form is expressed as follows:
\begin{equation}
\label{deqn_19}
f_{v}\left(s_{i}(t)\right)=\frac{v_{\max }}{2}\left(1-\cos \left(\pi \frac{s_{i}(t)-s_{\min }}{s_{\max }-s_{\min }}\right)\right),
\end{equation}
with $ v_{\max } $ denoting the maximum velocity. According to~\eqref{deqn_17}-\eqref{deqn_19}, the HDV model transfer function can be obtained as:
\begin{equation}
\label{deqn_20}	
G_{\text{HDV}}(\mathscr{S})=\frac{\beta \mathscr{S}+\alpha \dot{V}\left(s^{*}\right)}{\mathscr{S}^{2}+\left(\beta+\alpha\right) \mathscr{S}+\alpha \dot{V}\left(s^{*}\right)}.
\end{equation}

\vspace{0.2em}
\noindent
\textbf{Independent CAVs:} For the independent CAVs, we apply the following CACC (Cooperative Adaptive Cruise Control) controller to describe their car-following behavior~\cite{Milanes2014}:
\begin{equation}
\label{deqn_21}	
\begin{cases}
e_i(t)=s_i(t)-s_{\rm{0}}-t_{\rm{h}} v_i(t), \\
v_i(t)=v_i(t-\Delta t)+k_{\rm{p}} e_i(t)+k_{\rm{d}} \dot{e}_i(t),
\end{cases}
\end{equation}	
where $e_i(t)$ denotes the difference between the actual space and the desired space, with $\dot{e}_i(t)$ denoting its differential. In addition, $ s_{0} $ is the minimum safe space, $ t_{\rm{h}} $ is the desired time headway, $ \Delta t $ denotes the sampling interval, and $ k_{\rm{p}}, k_{\rm{d}} $ denote the control gains. Taking the acceleration $\dot{v}_{i}(t)$ as the control decision variable, the CAV model can be expressed as follows:
\begin{equation}
\label{deqn_22}	
\dot{v}_{i}(t)
=\frac{k_{\rm{p}}\left(s_i(t)-s_{0}-t_{\rm{h}} v_{i}(t)\right)+k_{\rm{d}} \left(v_{i-1}(t)-v_{i}(t)\right)}{k_{\rm{d}} t_{\rm{h}}+\Delta t}.
\end{equation}
Then, the CAV model transfer function is obtained as:
\begin{equation}
\label{deqn_23}	
G_{\text{CAV}}(\mathscr{S})=\frac{\frac{k_{\rm{d}}}{k_{\rm{d}} t_{\rm{h}}+\Delta t} \mathscr{S}+\frac{k_{\rm{p}}}{k_{\rm{d}} t_{\rm{h}}+\Delta t} }{\mathscr{S}^{2}+\left(\frac{k_{\rm{d}}}{k_{\rm{d}} t_{\rm{h}}+\Delta t}+\frac{k_{\rm{p}} t_{\rm{h}}}{k_{\rm{d}} t_{\rm{h}}+\Delta t}\right) \mathscr{S}+\frac{k_{\rm{p}}}{k_{\rm{d}} t_{\rm{h}}+\Delta t} }.
\end{equation}

\vspace{0.2em}
\noindent
\textbf{Mixed Platoons:} For a fair comparison between the MPF and MSL topologies in the mixed platoons, we utilize the standard linear quadratic regulator (LQR) method to design the control inputs based on the system dynamics~\eqref{deqn_10}. The performance index is characterized by the cost function $J$, given by:
\begin{equation}
\label{deqn_24}
J=\int_{0}^{\infty}\left(x(t)^\top Q x(t)+u_i(t)^\top R u_i(t)\right) dt,
\end{equation}	
where $Q=\operatorname{diag}\left(Q_{1}, \cdots, Q_{j}, \cdots, Q_{2m}\right) \in \mathbb{R}^{2m \times 2m},R \in  \mathbb{R}$ denote the penalty matrices for the system states and control inputs, respectively. Here, Bryson's Rule~\cite{Bryson1993} is employed to determine the parameters in both $ Q $ and $ R$. 
After determining the value of $Q$ and $R$, they are substituted into the algebraic Riccati equation:
\begin{equation}
\label{deqn_27}
A^\top P+P A-P B R^{-1} B^\top P+Q=0,
\end{equation}	
to solve the definite positive symmetric matrix $P$, and then apply $P$ to calculate the feedback gain $K$ with the equation:
\begin{equation}
\label{deqn_28}
K = R^{-1} B^\top P. 
\end{equation}	
Then, combined with~\eqref{deqn_10}, the mixed platoon model transfer function as:
\begin{equation}
\label{deqn_29}	
G_{m}(\mathscr{S})=C\left( \mathscr{S} I-\left( A+BK\right)\right)^{-1}H.
\end{equation}

\subsection{Numerical Results of String Stability}

In this subsection, we utilize the method in~\eqref{Eq:StringStability} to numerically analyze the performance of MPF topology, MSL topology, and pure CACC (\emph{i.e.}, all CAVs use the CACC model~\eqref{deqn_21}) in mixed traffic flow. Besides focusing on IFT and MPS, we also incorporate the role of CAV penetration rates to provide a more in-depth and comprehensive understanding of the impact of CAVs on mixed traffic. 

\begin{figure}[t]
\centering
\vspace{0.2cm}
\subfigure[$M = 4$]{\includegraphics[width=8.4cm]{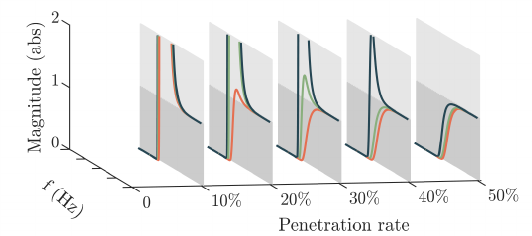}}\\
\vspace{-0.2cm}
\subfigure[$M = 6$]{\includegraphics[width=8.4cm]{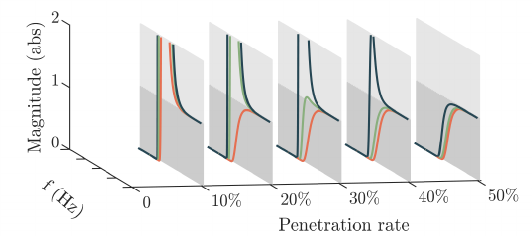}}\\
\vspace{-0.2cm}
\subfigure[$M = 8$]{\includegraphics[width=8.4cm]{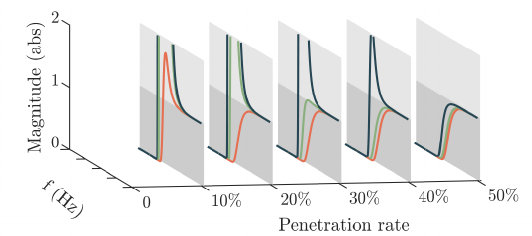}}\\
\vspace{-0.2cm}
\subfigure{\includegraphics[width=7cm]{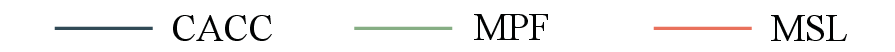}}
\vspace{-0.2cm}
\caption{Bode diagrams of mixed traffic under CACC, MPF topology, and MSL topology. The dark gray background and the light gray background represent the string stability region and the string instability region, respectively. The maximum platoon size $M$ is $4$, $6$, and $8$ in (a), (b), and (c), respectively. }
\label{fig:2}
\vspace{-0.5cm}
\end{figure}

We set total number of vehicles as $N=1000$, and consider three different values for the MPS  $M=4, 6, 8$ with reference to \cite{yao2023analysis,Zhou2021}. Then the string stability of mixed traffic at different MPS values is shown in Fig.~\ref{fig:2}. According to the string stability criterion~\eqref{Eq:StringStability}, the frequency domain in the Bode diagrams is divided into two phases according to magnitude, with magnitude less than or equal one denoting string stability and magnitude greater than one denoting string instability.

Different profiles in Fig.~\ref{fig:2} represent the influence of different kinds of IFT on string stability. It is evident that the MSL topology outperforms both the MPF topology and CACC in terms of effectively stabilizing traffic flow across various MPS values and penetration rates. Specifically, at a MPS of $4$, MSL topology achieves string stability with a penetration rate of $\SI{30}{\%}$, whereas MPF topology and CACC require a higher penetration rates of $\SI{40}{\%}$ and $\SI{50}{\%}$, respectively (see Fig.~\ref{fig:2}(a)). Further, with a higher MPS of $6$ or $8$, MSL allows for string stability at only a penetration rate of $\SI{20}{\%}$, which is still lower than the other topologies (see Fig.~\ref{fig:2}(b)(c)).  These observations reveal that ``looking behind" could further enhance the capability of CAVs in mitigating traffic perturbations in low penetration rates. 

In addition, Fig.~\ref{fig:2} also displays the effect of MPS on string stability. 
It is obvious that as the MPS increases, the string stability of mixed traffic at different IFT setups all obtains a significant improvement. The magnitude gets lower from Fig.~\ref{fig:2}(a) to Fig.~\ref{fig:2}(c), representing a better mitigation performance against perturbations when one increases the mixed platoon size, \emph{i.e.}, incorporating more neighbouring HDVs into the control design of the CAVs.  

\section{Nonlinear Traffic Simulations}
\label{Sec:4}
In this section, we present the results from the nonlinear traffic simulations. The specific models for different types of vehicles remain the same as those in Section~\ref{Sec:3}. Unlike the the previous string stability analysis, which is based on the linearized dynamics, the nonlinear OVM model~\eqref{deqn_17} is employed for HDVs in the following simulations.

\subsection{Simulation Setup}

In the simulations, the number of all the vehicles $N$ is $100$. The vehicle length is $ \SI{5}{m}$, and the maximum acceleration and deceleration of the vehicle are set to $\pm\SI{2}{m/s^2}$. The parameter setup is listed in Table~\ref{tab:1}. 
Note that the location of the CAVs are generated randomly in mixed traffic flow at different penetration rates. 

\begin{table}[htbp]
	\centering
    \vspace{-0.2cm}
	\caption{Parameter Setup in Simulations}
	\label{tab:1}
	\resizebox{\linewidth}{!}{
		\begin{tabular}{ccccccc}
			\toprule
			Parameter& $\alpha$& $\beta$& $s_{\min}$& $s_{\max}$& $v_{\max}$& $\Delta t$\\
			\midrule
			Value& $\SI{0.6}{/s}$& $\SI{0.9}{/s}$ & $\SI{2}{m}$ & $\SI{32}{m}$& $\SI{30}{m}$& $\SI{0.1}{s}$\\
			\bottomrule
			Parameter& $t_{\rm{h}}$& $s_{0}$& $k_{\rm{p}}$& $k_{\rm{d}}$& $Q_{i}$& $R$\\
			\midrule
			Value& $ \SI{1.0}{s} $ & $\SI{2}{m}$& $\SI{0.45}{/s}$& $\SI{0.25}{}$& $\SI{1}{}$& $\SI{1}{}$\\
			\bottomrule
	\end{tabular}}
\end{table}

To investigate the propagation of perturbations, we impose a perturbation on the head vehicle. At $t = \SI{10}{s}$, the head vehicle decelerates and then accelerates with a same intensity $ \SI{1}{m/s^2} $ for a same time duration $\SI{3}{s}$. Precisely, the velocity profile of the head vehicle is designed as follows:
\begin{equation}
\label{deqn_30}
v=
\begin{cases}
v^{*} & (0 \mathrm{~s}<t<10 \mathrm{~s}) \\
v^{*}-t+10 & (10 \mathrm{~s}<t<13 \mathrm{~s}) \\
v^{*}+t-16 & (13 \mathrm{~s}<t<16 \mathrm{~s}) \\
v^{*} & (16 \mathrm{~s}<t<T \mathrm{~s})
\end{cases},
\end{equation}	
where $v^{*}=\SI{15}{m/s}$, and the total simulation time is set to $T = \SI{150}{s}$. Then, we conduct simulations under different MPS, IFT, and CAV penetration rates. 

\subsection{Simulation Result}

\begin{figure*}[htb]
	\centering
    \vspace{0.1cm}
	\subfigure[CACC with $\SI{10}{\%}$ penetration rate]{\includegraphics[width=8cm]{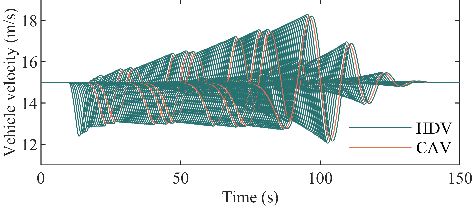}}
	\hspace{0.5cm}
	\subfigure[CACC with $\SI{20}{\%}$ penetration rate]{\includegraphics[width=8cm]{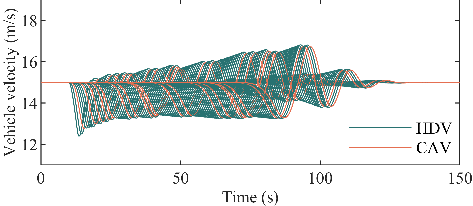}}\\
    \vspace{-0.2cm}
	\subfigure[MPF topology with $\SI{10}{\%}$ penetration rate]{\includegraphics[width=8cm]{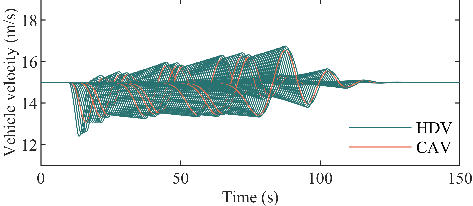}}
	\hspace{0.5cm}
	\subfigure[MPF topology with $\SI{20}{\%}$ penetration rate]{\includegraphics[width=8cm]{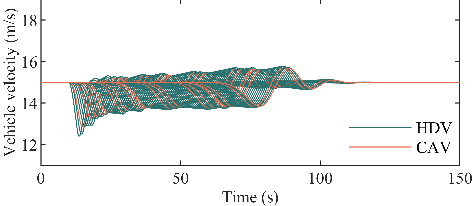}}\\
    \vspace{-0.2cm}
    \subfigure[MSL topology with $\SI{10}{\%}$ penetration rate]{\includegraphics[width=8cm]{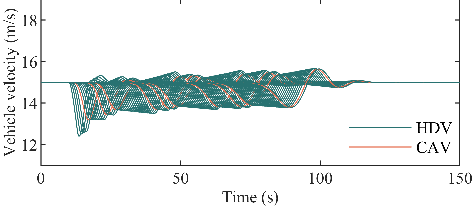}}
	\hspace{0.5cm}
	\subfigure[MSL topology with $\SI{20}{\%}$ penetration rate]{\includegraphics[width=8cm]{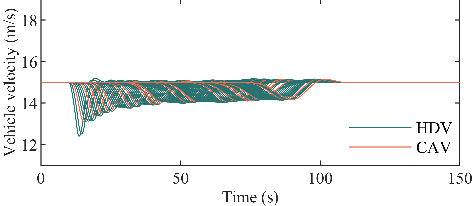}}\\
    \vspace{-0.2cm}
	\caption{Mixed traffic simulation results under CACC, MPF topology, and MSL topology with maximum platoon size $ M=6 $.}
	\label{fig:3}
	\vspace{-0.5cm}
\end{figure*}

Fig.~\ref{fig:3} demonstrates the propagation of velocity perturbations in mixed traffic flow under different IFT setups given a CAV penetration rate of $\SI{10}{\%}$ or $\SI{20}{\%}$. The results reveal that when the penetration rate is $\SI{10}{\%}$, the MSL topology, MPF topology, and CACC fail to ensure string stability. By contrast, when the penetration rate is $\SI{20}{\%}$, the MSL topology ensures string stability, while the MPF topology and CACC still fail to provide such assurance. These findings support the outcomes presented in Fig.\ref{fig:2} of Section~\ref{Sec:3}. Furthermore, upon comparing the results of Fig.\ref{fig:3}(a), (c), and (e) or Fig.\ref{fig:3}(b), (d), and (f), it becomes evident that both MSL and MPF exhibit superior abilities in enhancing string stability compared to CACC, with MSL demonstrating greater effectiveness. Due to page limit, figures of different topologies at other penetration rates and MPS setups are not included, but corresponding numerical values are provided to indicate the observed effects in the following discussions.

\begin{figure}[!tb]
	\centering
    \vspace{0.1cm}
	\subfigure[SD index comparison]{\includegraphics[width=8.5cm]{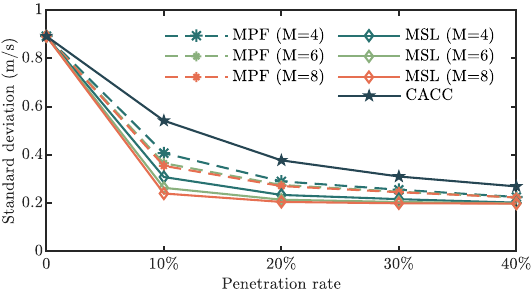}}\\
	\vspace{-0.3cm}
	\subfigure[MAD index comparison]{\includegraphics[width=8.5cm]{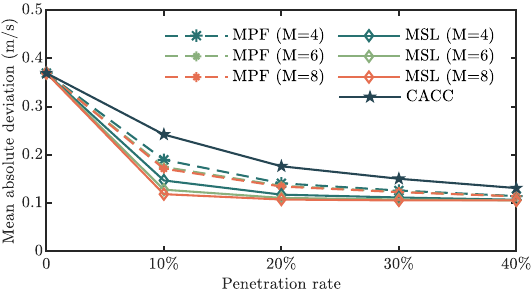}}\\
	\vspace{-0.2cm}
	\caption{Performance indexes under different topologies, different maximum platoon sizes and different penetration rates.}
	\label{fig:4}
	\vspace{-0.6cm}
\end{figure}

In order to quantitatively compare the performance of nonlinear mixed traffic under different setups of IFT, MPS, and CAV penetration rate, we select the velocity standard deviation (SD) and mean absolute deviation (MAD) as the performance indexes, defined as:
\begin{equation}
\label{deqn_31}
\text { SD }=\sqrt{\frac{1}{N-1} \sum_{t=0}^{T} \sum_{i=1}^{N}\left(v_{i}(t)-\bar{v}(t)\right)^{2}},
\end{equation}	
\begin{equation}
\label{deqn_32}
\text { MAD }=\frac{1}{N} \sum_{t=0}^{T} \sum_{i=1}^{N}\left|v_{i}(t)-\bar{v}(t)\right|,
\end{equation}
where $\bar{v}(t)$ is the average velocity at time $t$. Higher values of SD and MAD indicate larger velocity oscillations, and weaker traffic stability. We conduct simulations at various values of MPS and penetration rates, and the values for the two indexes of MPF, MSL and pure CACC are demonstrated in Fig.~\ref{fig:4}.

As can be clearly observed, the SD and MAD values all show a decreasing trend with the increase of the CAV penetration rate. Regarding the IFT, both MSL and MPF topologies have a stronger wave-dampening than pure CACC at an arbitrary CAV penetration rate. Furthermore, at different setups of MPS, MSL shows a better performance than MPF in attenuating perturbations. In addition, when MPS is increased from $4$ to $8$, the SD and MAD decrease under the perturbation effect. It illustrates that the stabilization effect of both MPF and MSL topologies increases with the expansion of platoon size, which is consistent with our analytical solutions in Section~\ref{Sec:3}. Moreover, when MPS reaches $6$, continuing to enlarge MPS for the MPF topology, \ie, incorporating more HDVs into the control consideration of the CAVs, does not bring significantly improvement on the traffic flow stability.

\section{Conclusions}
\label{Sec:5}
This paper presents a first comparison study between ``looking ahead" and ``looking behind" topologies for CAVs in mixed traffic. Particularly, two kinds of IFT, called MPF and MSL, are under consideration. Our method integrates both information flow topology and maximum platoon size to comprehensively evaluate the string stability of mixed traffic. The results show that both topologies outperform the conventional CACC in terms of string stability, and notably, MSL (``looking behind") demonstrates a greater advantage than MPF (``looking ahead"). Furthermore, a larger platoon size contributes to enhanced string stability of mixed traffic, but for the MPF topology, the extent of improvement diminishes as the platoon size exceeds a certain threshold. These results fill the gap of insufficient knowledge about 
the selection of information flow topologies and maximum platoon size configurations for mixed platoons.

For future research, one interesting direction is to compare the information flow topologies at a different control framework, such as structured control~\cite{wang2020controllability} or robust control~\cite{mousavi2022synthesis}. In addition, analyzing the effects of communication delay and HDV reaction time is also worth future investigation.

\bibliographystyle{IEEEtran}
\bibliography{IEEEabrv,Reference}

\end{document}

%% file: defs.tex
\def\0{{\bf 0}}
\def\1{{\bf 1}}

\def\eg{{\em e.g.}}
\def\ie{{\em i.e.}}

